\documentclass[conference]{IEEEtran}
\IEEEoverridecommandlockouts

\usepackage{amsmath,amssymb,amsfonts}
\usepackage{hyperref}
\usepackage{algorithmic}
\usepackage{graphicx}
\usepackage{textcomp}
\usepackage{xcolor}
\usepackage{csquotes}
\usepackage[inline,shortlabels]{enumitem}
\usepackage{listings}
\usepackage{booktabs}
\usepackage{colortbl}
\usepackage{subcaption}
\usepackage{soul}
\usepackage[capitalise,noabbrev]{cleveref}

\crefname{paragraph}{Section}{Sections}

\newcommand{\cpp}{C\texttt{++}}

\DeclareRobustCommand{\hlred}[1]{{\sethlcolor{red!50}\hl{#1}}}
\DeclareRobustCommand{\hlgray}[1]{{\sethlcolor{gray!30}\hl{#1}}}

\newenvironment{enumi}{\begin{enumerate*}[label=(\roman*)]}{\end{enumerate*}}

\def\BibTeX{{\rm B\kern-.05em{\sc i\kern-.025em b}\kern-.08em
    T\kern-.1667em\lower.7ex\hbox{E}\kern-.125emX}}
\begin{document}

\title{Dynamic Contract Analysis for Parallel~Programming~Models
\thanks{The authors would like to thank the German Federal Ministry of Research, Technology and
Space and the German federal states (http://www.nhr-verein.de/en/our-partners) for
supporting this work/project as part of the National High-Performance Computing (NHR) joint
funding program.}
}

\author{\IEEEauthorblockN{Yussur Mustafa Oraji}
\IEEEauthorblockA{\textit{Institute for Scientific Computing} \\
\textit{Technische Unversität Darmstadt}\\
Darmstadt, Germany \\
0009-0004-9922-3112}
\and
\IEEEauthorblockN{Alexander Hück}
\IEEEauthorblockA{\textit{Institute for Scientific Computing} \\
\textit{Technische Unversität Darmstadt}\\
Darmstadt, Germany \\
0000-0003-1931-773X}
\and
\IEEEauthorblockN{Christian Bischof}
\IEEEauthorblockA{\textit{Institute for Scientific Computing} \\
\textit{Technische Unversität Darmstadt}\\
Darmstadt, Germany \\
0000-0003-2711-3032}
}

\maketitle

\begin{abstract}
Parallel programming in high-performance computing depends on low-level APIs such as MPI, requiring users to manage synchronization and resources manually. 
Several correctness checking tools exist to help bug-free code development, though most target a single programming model, limiting their applicability.
Our previous work, the static analysis tool CoVer,
leverages a contract-based approach enabling users to specify custom error-checking rules and support emerging or unconventional programming models without requiring extensive new tooling.
However, static analysis cannot fully reason about runtime-dependent behavior such as pointer aliasing or indirect control flow.
To address this, we present CoVer-Dynamic, a dynamic analysis extension that reuses CoVer's contract language to provide a unified static-dynamic verification framework.
By enforcing the same contracts at runtime, CoVer-Dynamic improves classification accuracy and eliminates false positives on standardized MPI and OpenSHMEM benchmarks, while detecting errors beyond static analysis only.
Our evaluation shows that CoVer-Dynamic consistently outperforms the state-of-the-art correctness checker MUST, averaging a 2x speedup.
Finally, our results show limitations in the expressiveness of the contract language, motivating future work to support more error classes.

\end{abstract}


\section{Introduction}
\label{sec:introduction}


In high-performance computing (HPC) languages such as Fortran, C, and \cpp{} remain standard.
Within these environments, distributed-memory computations are managed via low-level APIs like MPI~\cite{messagepassinginterfaceforumMPIMessagePassingInterface2023},
OpenSHMEM~\cite{openshmemcommitteeOpenSHMEMApplicationProgramming2020}, and NCCL~\cite{nvidiaNVIDIACollectiveCommunication2025}. 
While this combination offers performance, the burden of correctly managing synchronization, resource lifetimes, and memory usage is placed on the developer.
Any usage violation can cause silent data corruption, deadlocks or program crashes.

To alleviate these issues, several correctness checking tools have been developed.
These can be categorized as primarily dynamic tools~\cite{hilbrichMUSTScalableApproach2010,schwitanskiRMASanitizerGeneralizedRuntime2024},
performing analysis at runtime, or static tools, analyzing only the source code~\cite{orajiVerifyingMPIAPI2026,burakSPMDIRUnifying2025,drosteMPIcheckerStaticAnalysis2015}.

Static tools typically have a global source-code view, enabling the detection of, e.g, data races existing in execution paths that are only triggered by specific test inputs at runtime, see \cref{fig:data_race_dynamic}. 
In contrast, while dynamic tools are limited to analysis of specific executed code paths, they can typically inspect runtime-dependent properties such as memory accesses and buffer overlaps for, e.g., data race detection, see \cref{fig:data_race_static}.

\begin{figure}
    \begin{subfigure}{0.49\columnwidth}
    \begin{lstlisting}
bool option = input();
^*\textbf{MPI\_Irecv(buf, \dots, req);}*^
if (option)
    ^*\textbf{buf[0]++;}*^
MPI_Wait(&req, ^*\dots*^);
        \end{lstlisting}
        \caption{Data race triggered if the input-based \texttt{option} is true.}
        \label{fig:data_race_dynamic}
    \end{subfigure}
        \hfill
    \begin{subfigure}{0.49\columnwidth}
        \begin{lstlisting}
int *buf1, *buf2;
^*\textcolor{gray}{// buffer setup omitted}*^
^*\textbf{MPI\_Irecv(buf1, \dots, req);}*^
^*\textbf{buf2[0]++;}*^
MPI_Wait(&req, ^*\dots*^);
        \end{lstlisting}
        \caption{Data race triggered if \texttt{buf1}, \texttt{buf2} overlap}
        \label{fig:data_race_static}
    \end{subfigure}
    \caption{Examples of data races difficult for automatic detection}
    \label{fig:data_race_examples}
\end{figure}

In addition, the aforementioned correctness tools typically target a single parallel programming model, for example MPI-focused checkers such as~\cite{hilbrichMUSTScalableApproach2010,drosteMPIcheckerStaticAnalysis2015}.
Extending such tools to support emerging models can therefore impose a substantial implementation burden.
Although there are efforts to provide a unified abstraction for (static) analysis~\cite{burakSPMDIRUnifying2025,burakExtendingSPMDIR2026},
these approaches still rely on hard-coded translation layers for a limited set of parallel models.

To address this model specificity, we previously introduced CoVer~\cite{orajiVerifyingMPIAPI2026}, a tool designed to decouple analysis logic and particular usage requirements of a parallel model. 
CoVer utilizes a language- and model-agnostic contract system in which correctness requirements are specified declaratively.
Instead of hard-coding any API usage rule for, e.g., MPI or OpenSHMEM into the analyzer, CoVer defines correctness requirements via a contract language specification based on string-based API annotations.
This allows for correctness checking of multiple parallel models without being dependent on their particular API.
However, since CoVer is a static checker, it remains inherently limited by its inability to resolve, e.g., pointer aliasing and other runtime-dependent conditions. 

To resolve these limitations, we present CoVer-Dynamic, which extends the applicability of the CoVer framework into the realm of dynamic analysis.
By reusing the same contract definitions for both static and dynamic contexts, we create a unified verification framework.
Therefore, CoVer-Dynamic has the advantage of dynamic tooling while retaining the model-agnostic contracts of the original CoVer tool.

Furthermore, with the shared contract base, we facilitate a \emph{hybrid} workflow to reduce runtime impact. 
Here, static analysis is first used to identify potential error sites, followed by targeted runtime instrumentation to validate these issues dynamically.
Unlike our prior assessed static-dynamic tool combinations~\cite{orajiCouplingStaticDynamic2025}, CoVer-Dynamic avoids inconsistencies caused by mismatched parallel model API or error-class support by maintaining a unified contract specification across both static and dynamic analyses.
Additionally, while the hybrid analysis was previously limited to data race detection only,
the shared understanding of contracts allows using the coupled analysis approach for all error classes.

In summary, our main contributions are:
\begin{itemize}
    \item A dynamic contract-based approach to error detection for parallel programming models called Cover-Dynamic.
    \item Support for the existing contract language of the static CoVer tool, allowing full reuse of these contracts.
    \item Seamless interoperability with the original CoVer implementation for targeted, low-overhead instrumentation.
\end{itemize}


The rest of this paper is structured as follows.
\cref{sec:background} introduces common errors in parallel models and describes the original Cover implementation and it's contract language.
\cref{sec:implementation} describes Cover-Dynamic and particular challenges of re-using static contracts during dynamic execution.
In \cref{sec:evaluation}, we evaluate our tool against the state-of-the-art MPI correctness checker MUST~\cite{hilbrichMUSTScalableApproach2010}. 
In particular, we also highlight that our hybrid workflow of pruning necessary instrumentation of dynamic analysis via static information can significantly improve runtime.
\cref{sec:related_work} briefly discusses related work.
Finally, we conclude in \cref{sec:conclusion}.

\section{Background}
\label{sec:background}

CoVer-Dynamic reuses the contracts of CoVer.
Thus, understanding the analysis approach of CoVer is necessary to discuss
the implementation of CoVer-Dynamic.
This section introduces some of the error classes representable within CoVer contracts,
as well as the methods used in CoVer to detect these issues in parallel code.

\subsection{Error Classes}
\label{sec:error_classes}

While parallel programming models offer a wide variety
of communication paradigms such as point-to-point, collective and remote memory access communication,
developing code using these communication methods often leads to subtle errors. 
This section introduces some of these errors using the example of MPI and OpenSHMEM.
The errors presented here are the same as those used for the CoVer publication \cite{orajiVerifyingMPIAPI2026}.

\paragraph{Missing Initialization or Finalization}
\label{sec:missing_init_finalize}

Many library-based parallel programming models require an explicit initialization before performing communication,
and a finalization once execution is finished.
Omitting these often leads to program crashes.

\paragraph{Local Data Races}
\label{sec:local_data_races}

When using non-blocking communication, the API call concludes before the communication is performed.
Thus, it is necessary to avoid read or write accesses to that buffer until it is explicitly completed.
Otherwise, the communication and local memory access are unordered, and the buffer value and/or communicated value are undefined.
We refer to these kinds of errors as local data races.

\begin{figure}
    \begin{lstlisting}
int src = 4;
#if defined(USE_MPI)
    MPI_Isend(^*\hlgray{src}*^, req1);
#else
    shmem_int_put_nbi(global_buf, ^*\hlgray{src}*^);
#endif
^*\hlred{src = 42}*^; // Error: Data race!
    \end{lstlisting}
    \caption{Example of a local data race, as the source data is written to before the communication completes.}
    \label{fig:data_race_ppms}
\end{figure}
Examples for MPI and OpenSHMEM are given in \cref{fig:data_race_ppms}.
Both the MPI and OpenSHMEM routines send the data contained in \texttt{src} to a target process.
However, \texttt{src} is written to before completion of the communication is ensured.
Thus, the actual data sent is undefined.
Future code examples will be limited to MPI for simplicity.

\paragraph{Handle Lifecycle}
\label{sec:handle_lifecycle}

When interacting with a parallel programming model, some structures are often represented as opaque handles.
For example, MPI defines handles for datatypes, communicators, windows, and requests,
and OpenSHMEM uses team and context handles.

These handles are resources, and must be constructed and disposed of correctly.
Some handles also need a more involved setup or destruction process;
a custom MPI datatype for example must also be explicitly committed after being constructed before being used for communication.
Issues with the management of handles are grouped under the handle lifecycle error class.

\begin{figure}
    \begin{lstlisting}
MPI_Isend(buf, ^*\hlgray{req1}*^);
MPI_Isend(buf2, ^*\hlred{req1}*^); // Error: Request handle lost!
    \end{lstlisting}
    \caption{Example of a request leak. The first handle is lost due to the repeated call.}
    \label{fig:double_request_use_example}
\end{figure}
A more difficult to detect MPI request leak, a subclass of handle lifecycle errors, is shown in \cref{fig:double_request_use_example}.
MPI allocates a request for each communication call. 
Using the same request handle for two calls without an intervening completion overwrites the original request handle,
causing the underlying request to leak.

\paragraph{Common Remote Memory Access Errors}
\label{sec:remote_memory_access_errors}

Remote Memory Access (RMA) allows for direct, one-sided access of inter-process memory.
However, before using RMA communication, some setup must be performed.
This is especially true for MPI:
MPI requires an RMA window to be created for communication beforehand.
OpenSHMEM does not \emph{require} this, as global and static variables are always considered exposed for communication,
though an OpenSHMEM memory management routine must be used otherwise.
MPI additionally requires communication to be performed in an epoch, which can be created depending on the synchronization method used;
exactly one (mixtures are not allowed) of fence, lock-unlock or Post-Start-Complete-Wait (PSCW).

\begin{figure}
    \begin{lstlisting}
MPI_Win_fence(^*\hlgray{win}*^);
MPI_Win_lock_all(^*\hlred{win}*^); // Error: Mixed Sync!
MPI_Get(&buf, win);
    \end{lstlisting}
    \caption{Example of a mixed synchronization error in MPI RMA (fence+lock); only one synchronization method may be used at one time per window.}
    \label{fig:mixed_sync_example}
\end{figure}
An example of this error class is given in \cref{fig:mixed_sync_example}.
Both fence and lock synchronization is used, which is not allowed and may cause undefined behavior.

\subsection{Verification using CoVer}
\label{sec:verification_using_cover}

CoVer \cite{orajiVerifyingMPIAPI2026} is a static correctness checker designed to verify code using parallel programming models.
It defines a contract language which allows specifying requirements on API functions,
which are then verified using an interprocedural data-flow analysis.

This section presents the core concepts of CoVer,
as these are important to understand for the introduction to CoVer-Dynamic.
The contract language especially is vital, as CoVer-Dynamic reuses the definitions.

\subsubsection{Contract Language}
\label{sec:contract_language}

\begin{figure}
    \begin{lstlisting}
int MPI_Finalize() CONTRACT( PRE { call!(MPI_Init) });

int MPI_Get(^*\dots*^) CONTRACT(
  PRE { call!(MPI_Init) }
  POST {
   no! (read!(*0)) until! (call_tag!(rma_complete,$:7)),
   no! (write!(*0)) until! (call_tag!(rma_complete,$:7))
  });

int MPI_Win_fence(int assert, MPI_Win win)
  CONTRACT( TAGS { rma_complete(1) });
int MPI_Win_unlock_all(MPI_Win win)
  CONTRACT( TAGS { rma_complete(0) });
    \end{lstlisting}
    \caption{Examples of contract definitions.}
    \label{fig:examples_contract_definitions}
\end{figure}
Consider the simple case of trying to detect a missing initialization error in MPI, i.e., not having called \texttt{MPI\_Init} (This is a simplified example not considering threading or MPI session support).
To encode this into a contract, it has to be reformulated into a requirement: When an MPI call is used, \texttt{MPI\_Init} \emph{must} have been called prior.

A contract encoding this requirement for \texttt{MPI\_Finalize} can be seen in \cref{fig:examples_contract_definitions}.
The function declaration of \texttt{MPI\_Finalize} is annotated with the new requirement.
\texttt{PRE} indicates that the requirement comes in effect prior to the call site of \texttt{MPI\_Finalize},
and \texttt{call!} is the operation that must be performed.

For a more complex example, consider detecting local data races.
The requirement is that \emph{after} each function call to a non-blocking communication,
\emph{no} reading/writing operations may occur \emph{until} the call is ensured to be completed using another API call.
This logical structure can be translated directly into the contract language similarly to the previous example.

However, this is not sufficient: It is not that \emph{any} reading or writing operations are forbidden, but only those to the communication buffer.
Additionally, not only must \textquote{some} completion call be made, but it must explicitly complete the communication,
using either the corresponding MPI request (if it was a point-to-point or collective call), or the MPI RMA window handle.

To correctly specify these restrictions, CoVer allows specifying whether parameters of the call sites must match a corresponding operation.
\cref{fig:examples_contract_definitions} provides an example of a requirement forbidding data races for \texttt{MPI\_Get}, while also requiring MPI initialization.
The latter check is done simply using the \texttt{PRE} block, while the former is done using two split requirements, one for forbidding reading and writing memory accesses respectively.
They are in the \texttt{POST} scope, indicating that the requirement is active after each call site of \texttt{MPI\_Get}.
For the memory accesses, the parameter \texttt{*0} is given, which indicates that explicitly accessing the memory pointed to by the first parameter is forbidden until the completion call,
which matches the MPI semantics, the first parameter of \texttt{MPI\_Get} being the communication buffer.

An MPI RMA call can be completed using multiple different mechanisms, for example through a fence or lock-unlock.
To encapsulate both, the corresponding functions are tagged with \texttt{rma\_complete}, and the parameter index housing the RMA window is stored into the tag.
The contract for \texttt{MPI\_Get} then uses a \texttt{call\_tag!} operation instead of \texttt{call!} with that tag as the target.
Thus, either function may complete communication as long as the window handle matches, which is ensured using the mapping from \texttt{\$} (the parameter in the tag) to the eighth parameter of \texttt{MPI\_Get},
which houses the window handle.

\begin{figure}
    \centering
    \includegraphics[width=0.8\columnwidth]{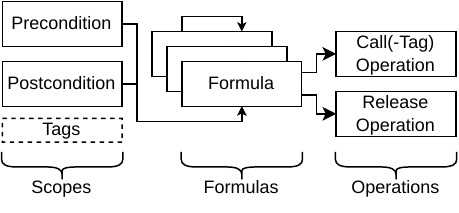}
    \caption{An overview of the contract language structure.}
    \label{fig:contr_lang_overview}
\end{figure}
The general contract structure can be seen in \cref{fig:contr_lang_overview}.
Contracts are strings attached to API functions, consisting at the highest level of \emph{scopes}.
The language defines three possible scopes: \texttt{PRE}, \texttt{POST} and \texttt{TAGS}.
\texttt{PRE} and \texttt{POST} may contain requirements that must hold before or after each call site of the API function,
while the \texttt{TAGS} scope can be used for \texttt{call\_tag!} operations as shown.
The requirements are specified using (possibly) nested formulas, with \emph{operations} being the atomic literals.
In the example for local data races, two requirements are given in a conjunction (i.e. both must be satisfied), though more complex formulas can be specified.

\subsubsection{Tool Architecture}
\label{sec:tool_architecture}

CoVer is implemented using compile-time passes on the LLVM Intermediate Representation \cite{lattnerLLVMCompilationFramework2004} (LLVM IR),
as well as a thin compile-time wrapper to enable analysis across translation units.
\begin{figure}
    \includegraphics[width=\columnwidth]{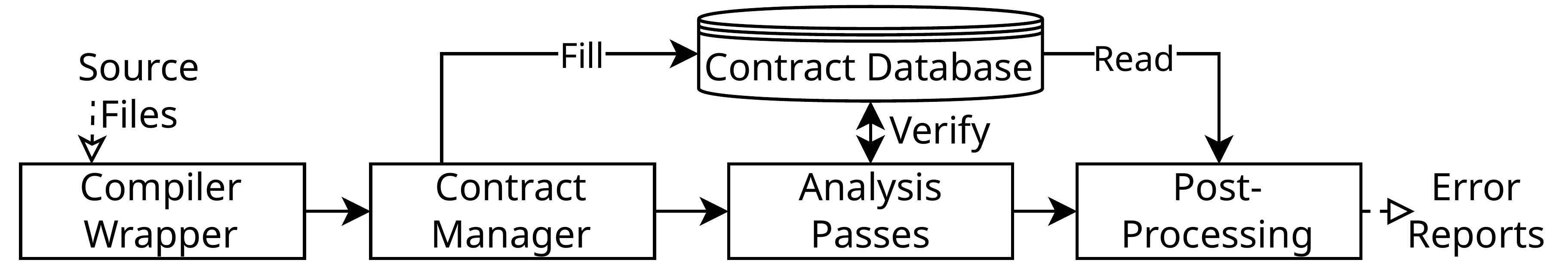}
    \caption{Overview of the CoVer architecture, from \cite{orajiVerifyingMPIAPI2026}}
    \label{fig:cover_static_architecture}
\end{figure}
The analysis process can be split into four parts:
\begin{enumi}
    \item invoking the compiler wrapper,
    \item the contract manager,
    \item running each analysis pass, and
    \item contract post-processing.
\end{enumi}
An overview is given in \cref{fig:cover_static_architecture}.

The contracts are given as string annotations using C/\cpp function declaration attributes.
To ease implementation, a macro which hides the specific attribute syntax is provided by the tool;
the \texttt{CONTRACT} macro as used in \cref{fig:examples_contract_definitions}.

The compiler wrapper intercepts all compilation steps,
and forces the underlying compiler to provide LLVM IR instead of generating object files.
Once the compilation reaches the link step, all IR files are linked together,
after which the wrapper runs the contract manager and analysis passes,
allowing the analysis to run across translation units.

The contract manager processes all contract annotations,
generating a \emph{contract tree} representation for each.
The contract tree is an object-based format which stores
the structure of each contract, i.e. the pre- and postconditions, the (nested) formulas and the operations they include.
Additionally, the types of the formulas are saved, which is one of
\begin{enumi}
    \item conjunctive, requiring all child formulas to be satisfied,
    \item disjunctive, requiring at least one child formula to be satisfied or
    \item exclusive or, requiring exactly one child formula to be satisfied.
\end{enumi}
All contracts, as well as a mapping between each function and the corresponding tags, are saved into a global \emph{contract database},
which is made accessible to the analysis passes.

CoVer includes three analysis passes, \texttt{PreCall}, \texttt{PostCall} and \texttt{Release}.
\texttt{PreCall} and \texttt{PostCall} check if a function was called before or after each callsite.
They are used if a call operation is present in the pre- or postcondition respectively.
\texttt{Release} is used for exclusively for release operations, which are always in the postcondition at the moment.
It checks whether the forbidden operation occurs before the releasing one (such as the data race check in \cref{fig:examples_contract_definitions}).
The passes do not run on the entire formulas. They depend solely on the atomic operations contained therein.
When they are finished, the result is written back to the contract database.

Now that the fulfillment status of all operations is known, a post-processing pass is able to resolve the formulas.
It recursively checks for the fulfillment of each nested formula up until the top level.
If the top-level formula is violated, an error is produced and reported to the user.

\section{Implementing dynamic contract analysis}
\label{sec:implementation}

We aim to create a new dynamic tool capable of checking the same contracts as the static tool CoVer \cite{orajiVerifyingMPIAPI2026}.
By reusing the contract definitions, existing definitions remain valid,
allowing the same checks for both tools while providing the benefits of dynamic analysis.

\subsection{Challenges of Dynamic Contract Analysis}
\label{sec:challenges_dynamic_contracts}

As CoVer's contracts were never intended to be used in a dynamic context,
this leads to some implementation challenges for our new approach.
This section discusses these challenges, namely the availability of the contract definitions,
the need for model-independent function call interception, and live error reporting.

\paragraph{Contract Availability}
One base issue is the accessibility of the contracts.
While these can be read statically from the annotations,
they become unavailable once the program is compiled.
Internally, the annotations are given as function declaration attributes,
which are only relevant to the compilation process and therefore discarded before runtime.
Thus, our dynamic tool needs an alternative way to read in the contracts.

\paragraph{Call Interception}
Additionally, to achieve model-agnostic analysis,
the new dynamic tool must be completely unaware of the programming model used.
Common points of interception for correctness checkers, such as the PMPI interface for MPI,
are not viable due to this restriction.

\paragraph{Live Error Reporting}
Finally, another obstacle is the structure of the contracts.
They are designed in formulas, and the static tool can resolve these in a post-processing step.
This is significantly harder dynamically: The program may have multiple exit points,
and a crash can cause all analysis data to be lost.
Since crashes are common in buggy code (the primary reason for using a correctness checker), the issue is particularly insidious, as a crash may be caused by the specific bug the tool is trying to detect.

While we could instead save analysis data to disk and perform a post-mortem analysis,
this would incur significant overhead (especially on parallel file systems used in clusters), and is therefore not feasible.

We have addressed each of these issues within our tool, the details of which will follow in the next sections.

\subsection{Compile-Time Instrumentation}
\label{sec:compile_time_instrumentation}

\begin{figure}
    \centering
    \includegraphics[width=0.8\columnwidth]{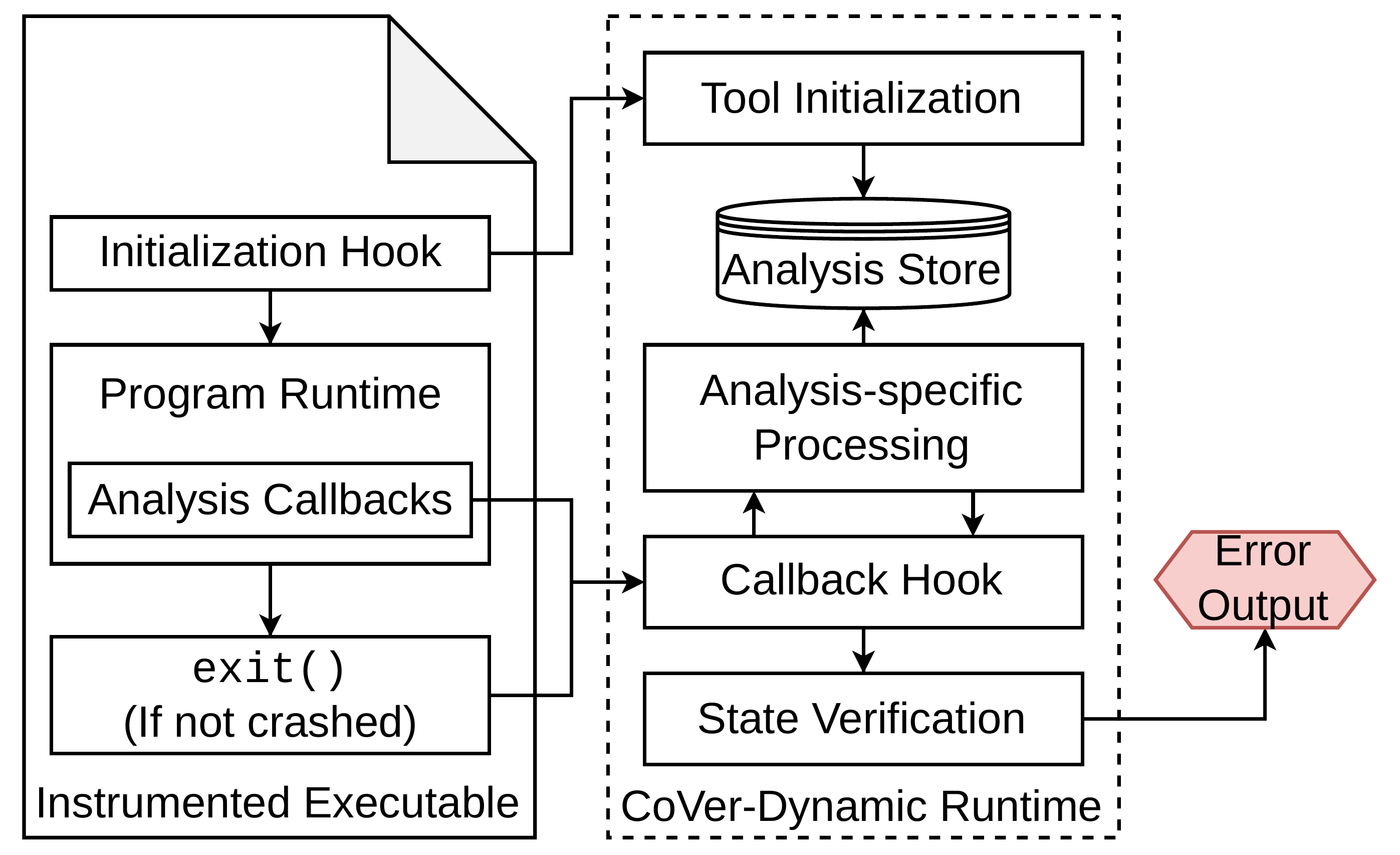}
    \caption{An overview of the dynamic analysis library at runtime.}
    \label{fig:runtime_overview}
\end{figure}
\begin{figure}
    \begin{lstlisting}
// Callback function declarations
void CB_Initialize(int* argc, char*** argv,
                      ContractDB_t* DB);

// ^*\textbf{\textcolor{green!50!black}{... -> Vararg}}*^: For every parameter in the callsite,
// first the parameter size, then the actual parameter
void CB_Function(bool isRel, void* function,
                 int num_params, ...);

void CB_Memory(bool isRel, void* buf, bool isWrite);
    \end{lstlisting}
    \caption{The callbacks to the dynamic tool generated by CoVer.}
    \label{fig:callback_definitions}
\end{figure}
As CoVer already parses the contracts from the source code,
we decided to implement compile-time instrumentation as an optional pass (using an additional option in its compiler wrapper),
which then adds the contract information to the executable.
The dynamic tool runtime itself is built as a static library, which is linked into the executable and performs the analyses using specially added callbacks.

An overview of the tool structure is given in \cref{fig:runtime_overview}, with the additional instrumented functions listed in \cref{fig:callback_definitions}.
However, the \texttt{isRel}, \texttt{argc} and \texttt{argv} arguments will only be referenced in \cref{sec:static_dynamic_coupling},
and are used to implement the coverage warnings for the static-dynamic coupled workflow as described in \cite{orajiCouplingStaticDynamic2025}.

\paragraph{Initialization of CoVer-Dynamic}
Once CoVer finishes parsing the contracts,
they are encoded into a C-style struct representation.
All contracts are packaged into a database,
and embedded into the executable as a global variable.

A callback to the dynamic tool is inserted at the start of the program with a pointer to the database (\texttt{CB\_Initialize} in \cref{fig:callback_definitions}).
The function receives a pointer to the global contract database struct, which was previously added to the executable using the static instrumentation,
which serves as the primary method for the dynamic tool to receive the instrumented contract information.

To do this, some additional processing is necessary.
For example, some operations may reference function names (e.g. \texttt{call!(MPI\_Init)}).
The dynamic tool cannot link against any programming model.
Thus, even if the name of the function is available,
the tool cannot check if the function is called, as it cannot identify it.
CoVer-Dynamic has no understanding of MPI, only the contract string.
During runtime, it has no ability to check if any called function is \texttt{MPI\_Init} as the name is no longer available,
only a function pointer remains.
We resolve this by not only storing the function name at compile time,
but also a pointer to the function in each representation of a call operation.
During runtime, it can then be used as an identifier of the corresponding function.

\paragraph{Intercepting API Calls}
To intercept the calls to the programming model, we decided to implement a callback system.
Every function mentioned in a contract, or has a contract attached, is added to a \textquote{relevant functions} list.
Then, for every callsite of the functions therein, a callback to the dynamic tool is created (\texttt{CB\_Function} in \cref{fig:callback_definitions}).

This callback function is a catch-all for all functions mentioned in a contract, as it is not possible to define different callbacks for different functions:
Due to the strict independence of the programming model and verification tool, constructing a specific function target for a callback ahead of the analysis is not possible.
Thus, as the callback must be a catch-all, to supply all needed information on the callsite it must also be variadic to support different numbers of arguments of the instrumented functions.

The first and second parameter to the callback are the pointer to the relevant function, and the number of parameters thereof.
Then, for every argument used in the callsite of the relevant function, the callback receives first the size of the parameter, and the parameter itself.
Specifying the size of each argument is critical, as otherwise the dynamic tool would have no way of knowing how many bytes to read for each argument.
This provides the dynamic tool with access to the argument values of each relevant function for analysis,
and functions referenced in the contracts can be compared to the pointer in the first parameter.

\paragraph{Additional Instrumentation}
Callbacks are also inserted for local memory accesses, storing the type (read/write) and the pointer to the modified memory location (\texttt{CB\_Memory} in \cref{fig:callback_definitions}).
This is necessary to reason about, for example, local data races, as otherwise the tool is unable to detect local memory accesses by the host program.
The callback is simple, with the parameters being the pointer to the memory accessed, and whether the access was writing or reading.

Finally, program exit is captured as well using the \texttt{atexit} function, which gives a callback once the program (implicitly) calls \text{exit}.
This is \emph{not} necessary for the error output, which as mentioned would prevent the tool from reporting errors if the program crashes,
but to capture the semantics of the \texttt{PostCall} operation.
It indicates that all \texttt{PostCall} operations that are still pending have been violated;
for example, if the contract \texttt{POST \{call!(MPI\_Finalize)\}} was not verified yet at program exit,
it is clear that the program has a missing finalization.

\subsection{Runtime Interaction}
\label{sec:runtime_interaction}

After the compile-time instrumentation inserts the contracts and callbacks,
we link the executable to a static library containing the definitions of the callback functions.
This is the core of the dynamic tool:
During runtime, this library performs the necessary checks to verify the annotated contracts.

\subsubsection{Managing Analysis Instances}
\label{sec:analysis_instances}

\begin{figure}
    \centering
    \includegraphics[width=\columnwidth]{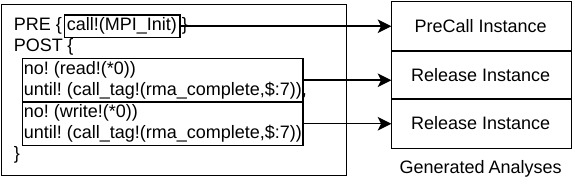}
    \caption{The created analysis instances for the contract given in \cref{fig:examples_contract_definitions}.}
    \label{fig:contract_analysis_instances}
\end{figure}
During the initialization callback, for every contract in the contract database, the contract formulas are recursively inspected for all occurrences of contract operations (e.g. \texttt{PreCall}, \texttt{PostCall}, \texttt{Release}).
An instance of a corresponding analysis object is created for each operation found, and then added to the \emph{analysis store}.
For the example contract in \cref{fig:examples_contract_definitions} attached to \texttt{MPI\_Get}, this can be seen in \cref{fig:contract_analysis_instances}.
The analysis instances are responsible for exactly one operation each.
Thus, for the example given in \cref{fig:contract_analysis_instances}, the given \texttt{PreCall} instance checks whether \texttt{MPI\_Init} was called prior to \texttt{MPI\_Get}.

The required information to perform this analysis is given through the instrumented callback functions.
All analyses which have not yet been resolved (neither violated nor fulfilled) are informed of every callback.
The analyses can then independently decide whether that information is relevant to them, and return whether in the current program state the analysis is fulfilled, violated, or (yet) unknown.

For example, a call to \texttt{MPI\_Send} is irrelevant to the \texttt{PreCall} operation given in \cref{fig:contract_analysis_instances},
but a call to \texttt{MPI\_Init} would fulfill it, and a call to \texttt{MPI\_Get} violate it.
While a call to \texttt{MPI\_Send} prior to \texttt{MPI\_Init} would also be erroneous, the compartmentalization of contracts and the analysis instances means
that a different analysis object is responsible for checking this error for \texttt{MPI\_Init}.

All analyses which return a resolved state (whether fulfilled or violated) are removed from the analysis store, which leads to them no longer receiving callbacks.
This improves the performance of the tool, by avoiding unnecessary high-cost function calls.
To further improve performance, analyses may directly state callback classes which are of no relevance to them.
For example, memory accesses are irrelevant for any \texttt{PreCall} operation, where only function calls are relevant.
Additionally, for \texttt{Release} operations memory accesses are only relevant if the forbidden operation therein is a memory access, such as local data races,
but not if it is a function call, for example when checking that no communication occurs using an MPI type that has yet to be committed.
This opts out the corresponding analyses from the memory callbacks entirely, reducing overhead.
Avoiding memory callbacks is especially important due to the sheer number of callbacks involved,
considering that \emph{all} memory accesses are being instrumented.
The need for optimization and effect of the instrumentation on runtime is further elaborated on in \cref{sec:overhead_analysis}.

\subsubsection{State Verification}
\label{sec:state_verification}

As described in \cref{sec:implementation}, it is necessary to perform error output immediately once a contract is violated.
However, this cannot be done by simply reporting an error once an operation is violated.
Contract formulas can be specified in (nested) conjunctions, \emph{disjunctions and exclusive-or relations},
meaning that one violated operation does not necessarily cause a violation in the entire contract.
\begin{figure}
    \begin{lstlisting}[language=]
PRE {
  call_tag!(epoch_fence_create,$:7) MSG "Fence epoch" ^*\textbf{\^}*^
  call_tag!(epoch_lock_create,$:7)  MSG "Lock epoch" ^*\textbf{\^}*^
  call_tag!(epoch_pscw_create,$:7)  MSG "PSCW epoch"
}
    \end{lstlisting}
    \caption{Example of a contract for \texttt{MPI\_Get} requiring an open epoch while forbidding mixed synchronization.}
    \label{fig:contract_mixed_sync}
\end{figure}
This can be seen when specifying a contract for mixed RMA synchronization (c.f. \cref{sec:error_classes}),
a simple example of which is shown in \cref{fig:contract_mixed_sync}.
The given contract is satisfied if exactly one of the call operations is fulfilled.
If none are, no epoch has been opened, and if multiple are, there has been mixed synchronization.
The static CoVer analysis resolves formulas during post-processing,
but in a dynamic tool a unified exit point is not guaranteed.
Thus, once a formula is violated it must be detected and reported immediately.

This is the task of the \emph{state verification} step seen in \cref{fig:runtime_overview}.
Any time an analysis is resolved, the library moves to state verification.
It is not sufficient to only perform this step for violated analysis instances,
as then exclusive-or formulas may fail to be invalidated;
multiple \textquote{fulfilled} contracts lead to a violation in the broader scope.

To give a short overview, the state verification step will always start at the level of an operation, which is an atomic formula.
Errors propagate upward throughout the formula tree, and, as long as they are not benign and captured by a disjunction or exclusive-or as irrelevant to wider correctness,
the top-level scope formula is inevitably marked violated.
At that point, the state of the program indicates a definite contract violation, and error output is performed.

In detail, once an analysis is fulfilled / violated, the tool moves to state verification.
If the type of the parent formula is a conjunction, and the current formula is violated, then the parent must necessarily be violated as well.
Thus, the tool will invalidate the parent and re-run the state verification using the parent formula.
For a disjunction, this will only occur if there are no siblings which have unknown state,
where the program has not run long enough to decide for violation or fulfillment.
For an exclusive-or, it checks whether multiple children have been fulfilled, or all of them violated.

Should the current formula already be a top-level formula (i.e. the entire pre- or postcondition), and it was violated,
it is guaranteed that an error has occurred.
The tool will then perform error output by recursively inspecting the children of the contract formula, reporting each violation.

Consider the example contract for mixed-sync detection given in \cref{fig:contract_mixed_sync}.
Should one child formula, such as the one responsible for fence, be violated, as the two siblings are of unknown state the full contract state is unknown as well.
If the formula for lock completion is fulfilled, the full contract stays in unknown state,
and a fulfillment of the last sibling would lead to a contract violation of the parent,
while a violation of it to a contract fulfillment thereof. 
As it is an exclusive-or, exactly one subformula should be fulfilled, not multiple.

\subsection{Static-Dynamic Coupling}
\label{sec:static_dynamic_coupling}

Dynamic tooling can cause significant runtime overhead and, also, may fail to report errors if specific code paths are not visited during execution, see~\cref{fig:data_race_dynamic}.
We can avoid these issues by using static-dynamic coupling, as shown in \cite{orajiCouplingStaticDynamic2025}:
\begin{itemize}
    \item Overhead can be significantly reduced using statically-guided filtering
    \item Unvisited locations can be checked for using a coverage-based approach
\end{itemize}

As static tools generally tend to favor false positives over false negatives,
the overhead induced by dynamic tooling can be reduced significantly by only instrumenting those locations where a static tool reports an error.

Furthermore, since static tools will inspect and report issues across the entire code,
the dynamic tool can check whether all locations that were reported by the static tool have actually been visited,
avoiding missing error reports due to unvisited locations containing errors.
For example, if the static tool detects the issue in \cref{fig:data_race_dynamic}, but the dynamic tool does not visit the relevant location,
a coverage error is reported.

\subsubsection{Filtered Instrumentation}
The filtered instrumentation can be easily implemented: As the instrumentation is performed just after a CoVer analysis using the compile wrapper,
the only required modification is selectively omitting the instrumentation when a code location is not referenced in an error report.
In the spirit of generality, CoVer-Dynamic can also read reports from other tools when the output is given using the exchange format defined in \cite{orajiCouplingStaticDynamic2025}.

\subsubsection{Coverage-Based Error Reports}
\begin{figure}
    \centering
    \includegraphics[width=0.9\columnwidth]{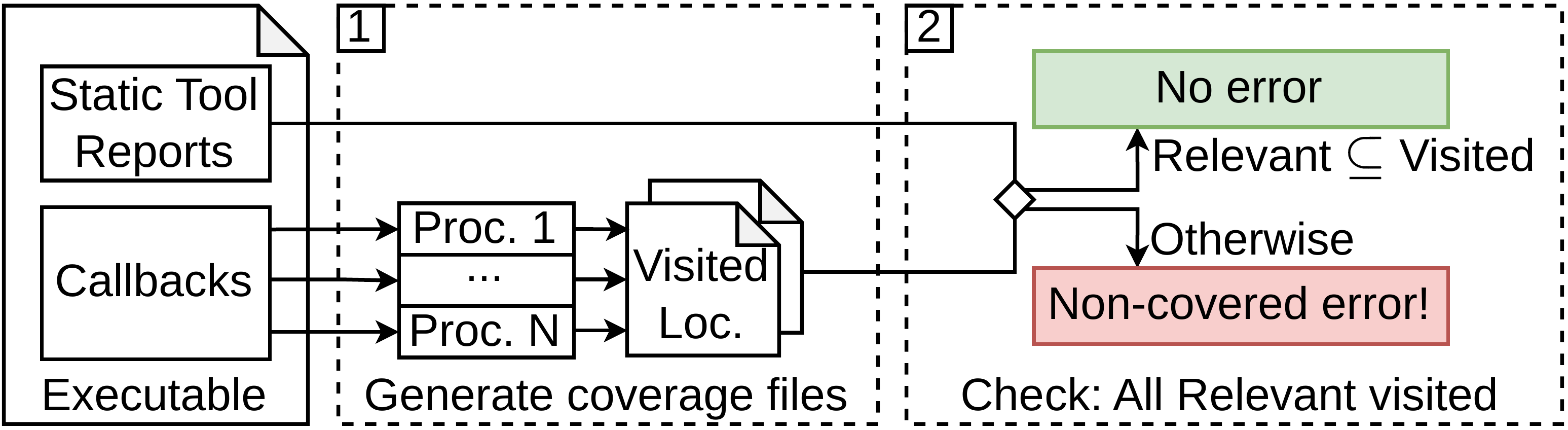}
    \caption{Coverage error generation process: (1) Normal execution incl. dumping of visited locations. (2) Visited locations are collected and compared to those mentioned in a static report.}
    \label{fig:static_dynamic}
\end{figure}
Implementing the coverage errors is slightly more involved; an overview of this implementation can be seen in \cref{fig:static_dynamic}.
First, CoVer-Dynamic must be made aware which locations are relevant (referenced in a static tool report), and which ones are not.
When using instrumentation filtering, this is implicit: No irrelevant instrumentation will be present.
However, while static tools are typically more prone to false positives than false negatives, they are still possible.
Thus, to ensure high accuracy the default remains full instrumentation.
Instead, we extended the instrumented callback functions with another parameter (given in \cref{fig:callback_definitions} as \texttt{isRel}),
which indicates whether the location of the callback was referenced by a static tool in an error report.
Using this information, CoVer-Dynamic can then collect all locations which were marked relevant at runtime.

Unfortunately, this by itself is insufficient:
Given an MPI code, where ranks 0 and 1 separately encounter one unique relevant location,
both ranks would independently report a coverage error for the corresponding location of the other process.
MUST collects the list of encountered locations across each process via MPI \cite{orajiCouplingStaticDynamic2025}.
However, due to the model-agnostic nature of CoVer-Dynamic, this is not possible here.

Instead, on program exit CoVer-Dynamic writes a unique file per process with a shared prefix,
containing the location of each visited relevant location (Step 1 in \cref{fig:static_dynamic}).

To generate the coverage information, the executable is started again (serially) using a special \texttt{--cover:check-coverage} flag (Step 2 in \cref{fig:static_dynamic}).
The program arguments are captured by CoVer-Dynamic at startup (see \cref{fig:callback_definitions}), which allows it to recognize the flag prior to the actual program execution.
CoVer-Dynamic then skips executing the actual program, instead collecting each visited location from the previously generated files.
As it is executing within the same executable, the embedded list of relevant locations is still valid,
making it possible to compare the actual locations visited and written to disk previously to the ones reported by the static tool.
Should these not match up, corresponding coverage errors are generated, after which the program exits.

Assume that in \cref{fig:data_race_dynamic} \texttt{option} is false for both process 1 and 2.
The data race will not materialize, but a static tool will report an issue in lines 2 (the MPI call) and 4 (the conflicting memory access).
Using CoVer-Dynamic, it will emit line 2 as visited, but not line 4.
Thus, when executing using the coverage check flag, this line will be missing in all coverage files, causing a coverage error to be emitted.
Conversely, if one rank does enter line 4, it will report a data race by itself, and emit line 4 as visited.
When collating the coverage info, all relevant locations are marked visited and no coverage error is reported, avoiding duplicate or invalid reports.

Here, a significant advantage of the contract-based approach becomes visible:
The original publication for static-dynamic coupling focused solely on local data race detection \cite{orajiCouplingStaticDynamic2025}.
But our implementation in CoVer-Dynamic is fully independent of the error class, instead requiring only a shared understanding of the contract definitions.
As this is always the case for CoVer and CoVer-Dynamic, it transparently extends the applicability of the coupled workflow to arbitrary error classes,
as long as they are representable using the contract language.

\section{Evaluation}
\label{sec:evaluation}

We performed various tests in order to verify the efficacy of the new dynamic contract analysis.
These tests are split into two parts:
\begin{enumi}
  \item classification quality tests using the existing RMARaceBench \cite{schwitanskiRMARaceBenchMicrobenchmarkSuite2023} and MPI-BugBench \cite{jammerMPIBugBenchFrameworkAssessing2025} (level 1) test suites, and
  \item overhead analysis using the proxy applications LULESH \cite{karlinLULESH20Updates2013}, PRK Stencil \cite{vanderwijngaartParallelResearchKernels2014} and TeaLeaf \cite{TeaLeaf2025}.
\end{enumi}
We compare CoVer-Dynamic with the state-of-the-art correctness checker MUST \cite{hilbrichMUSTScalableApproach2010} v1.11.2,
including the modifications for static-dynamic coupling introduced in \cite{orajiCouplingStaticDynamic2025}.

All tests were run on the Lichtenberg II cluster, using OpenMPI 5.0.5 for the classification quality tests and OpenMPI 4.1.6 for the overhead analysis.
This is due to an incompatibility of MUST with newer OpenMPI versions with a high number of MPI ranks, which caused deadlocks in our testing.
We used the Clang compiler (development) version 22 for both compiling the tests and tools.
The tool sources, scripts and results produced are available at \cite{orajiArtifactDynamicContract}.

\subsection{Classification Quality}
\label{sec:classification_quality}

All tests from MPI-BugBench level 1 and RMARaceBench, for a total of 458 tests, were analyzed using 5 tool configurations:
\begin{enumi}
    \item CoVer-Dynamic,
    \item CoVer,
    \item CoVer-filtered,
    \item MUST,
    \item MUST-filtered.
\end{enumi}
The \texttt{-filtered} suffix refers to dynamic analysis with targeted instrumentation,
where only those code locations are instrumented that may lead to an error as determined by a static analysis tool (here: CoVer).
This is part of the coupled analysis workflow, which improves runtime overhead at the cost of accuracy.

Once a test is executed, it is categorized as one of the following:
\begin{enumi}
  \item true positive (TP),
  \item true negative (TN),
  \item false positive (FP),
  \item false negative (FN),
  \item timeout (TO).
\end{enumi}
Additionally, two \emph{non-covered} classifications are used to represent cases arising from static-dynamic coupling (see \cref{sec:static_dynamic_coupling}).
When a dynamic tool recognizes that it did not visit a location that was marked as containing an error by the static tool,
it may report a coverage error.
This can improve accuracy by providing additional error reports.
CoVer, as a static tool, will not report coverage warnings. MUST only reports coverage warnings if instrumentation is filtered as well (the functionality is bound together).
For CoVer-Dynamic, coverage warnings can always be generated, no matter if instrumentation is filtered or not.
The non-covered classifications are categorized into NC-TP (not covered, true positive) and NC-FP (false positive).
Finally, the accuracy (A) is defined as the fraction of the positive results (TP, TN, NC-TP) over the total number of tests. 

\begin{table}
  \caption{Classification quality results.}
  \centering
  \setlength{\tabcolsep}{0.38em}
  \footnotesize
  \begin{tabular}{lrrrrrrrc}
  \toprule
  \multicolumn{9}{l}{\textbf{RMARaceBench} \scriptsize(MPI local data races)}\\
  \midrule
  & TP & TN & FP & FN & NC-TP & NC-FP & TO & A \\
  \midrule
  CoVer-Dynamic & 35 & \cellcolor{green!20}{31} & \cellcolor{green!20}{0} & 4 & 3 & 0 & 0 & \cellcolor{green!20}{0.95} \\
  CoVer & 34 & 28 & 3 & 8 & - & - & 0 & 0.85 \\
  CoVer-filtered & 32 & \cellcolor{green!20}{31} & \cellcolor{green!20}{0} & 7 & 3 & 0 & 0 & 0.90 \\
  MUST & \cellcolor{green!20}{39} & 27 & 1 & \cellcolor{green!20}{3} & - & - & \cellcolor{red!40}{3} & 0.94 \\
  MUST-filtered & 33 & 28 & 3 & 7 & 2 & 0 & 0 & 0.86 \\
  \midrule
  \addlinespace
  \midrule
  \multicolumn{9}{l}{\textbf{MPI-BugBench} \scriptsize(error classes from \ref{sec:error_classes})}\\
  \midrule
  & TP & TN & FP & FN & NC-TP & NC-FP & TO & A \\
  \midrule
  CoVer-Dynamic & 16 & \cellcolor{green!20}{23} & \cellcolor{green!20}{0} & 2 & 6 & 0 & 0 & 0.96 \\
  CoVer & \cellcolor{green!20}{24} & 22 & 1 & \cellcolor{green!20}{0} & - & - & 0 & \cellcolor{green!20}{0.98} \\
  CoVer-filtered & 16 & \cellcolor{green!20}{23} & \cellcolor{green!20}{0} & 2 & 6 & 0 & 0 & 0.96 \\
  MUST & 21 & 22 & 1 & 2 & - & - & \cellcolor{red!40}{1} & 0.93 \\
  MUST-filtered & 13 & 22 & 1 & 3 & 0 & 0 & \cellcolor{red!40}{8} & 0.90 \\
  \midrule
  \addlinespace
  \midrule
  \multicolumn{9}{l}{\textbf{RMARaceBench} \scriptsize{(OpenSHMEM tests)}} \\
  \midrule
  & TP & TN & FP & FN & NC-TP & NC-FP & TO & A \\
  \midrule
  CoVer-Dynamic & \cellcolor{green!20}{28} & \cellcolor{green!20}{24} & \cellcolor{green!20}{0} & \cellcolor{green!20}{5} & 2 & 0 & 0 & \cellcolor{green!20}{0.92} \\
  CoVer & \cellcolor{green!20}{28} & 22 & 2 & 7 & - & - & 0 & 0.85 \\
  CoVer-filtered & 25 & \cellcolor{green!20}{24} & \cellcolor{green!20}{0} & 8 & 2 & 0 & 0 & 0.86 \\
  \bottomrule
  \end{tabular}
  \vspace{0.5em}
\begin{flushleft}
\scriptsize
\hspace*{0.5em} TP\textuparrow: True Positive, TN\textuparrow: True Negative, FP\textdownarrow: False Positive, FN\textdownarrow: False Negative\\
\hspace*{0.5em} Of those, best results highlighted green.\\
\hspace*{0.5em} TO\textdownarrow: Timeout (30s, highlighted red), A\textuparrow: Accuracy\\
\hspace*{0.5em} \textuparrow: Higher is better, \textdownarrow: Lower is better\\
\smallskip
\hspace*{0.5em} Forwarded static tool errors due to dynamic tool coverage miss (see \cref{sec:static_dynamic_coupling}):\\
\hspace*{0.5em} NC-TP: Non-Covered True Positive, NC-FP: Non-Covered False Positive\\
\end{flushleft}
  \label{tab:results_corr}
\end{table}
The results of our analysis can be seen in \cref{tab:results_corr}.
However, it is important to note that the table excludes tests for remote RMA data races,
invalid parameters (e.g. using \texttt{NULL}, wrong MPI constant, etc.), data type mismatches, and deadlock detection.
These are currently not representable in the contract language defined by CoVer, and lead to trivial false negatives. 
Nevertheless, we executed these tests to ensure no false positives were reported (of which none occurred), though they are excluded from Table \ref{tab:results_corr} to avoid skewing the results with known functional gaps.
Extending support for these error classes is considered future work.

The classification quality results are generally favorable for CoVer-Dynamic.
The dynamic analysis achieved the highest total accuracy.
Additionally, no false positives were reported on any test.
This is the main advantage of dynamic analysis: the elimination of false positives compared to pure static analysis.
CoVer-Dynamic was also able to detect additional errors in contrast to its static counterpart.
For example, CoVer is unable to inspect function pointers, and thus cannot check the interprocedural control flow for these locations.
These are not an issue for CoVer-Dynamic, which is therefore able to correctly report errors for these paths.

MUST is capable of reporting more true positive errors than CoVer(-Dynamic).
This is due to the tight integration of MUST and MPI, allowing it to more deeply understand error behavior.
The excluded error classes due to the inability to represent them using the contract syntax are one example of this,
as well as the higher rate of true positives in the MPI-BugBench tests.
For example, these are due to tests containing data races with memory accesses in the middle of a buffer.
As CoVer does not understand how MPI determines the size of the buffer to send (as it is encoded using the MPI data type, number of elements, stride, etc.),
it cannot determine whether the memory access is conflicting, and avoids reporting an error.
As MUST has a built-in understanding of MPI call semantics, it can easily determine the conflicting access and report an error if needed.

However, the disadvantage of this tight integration is twofold, both stemming from the inherent \emph{dependence} which emerges from it.
MUST struggles with some tests, at times deadlocking and creating a timeout.
As MUST is deeply tied to MPI, it is very sensitive to the execution environment.
Changes such as different MPI versions or network transport can have significant impacts on the ability of the tool to function.

\begin{figure*}
  \centering
  \includegraphics[width=0.95\textwidth]{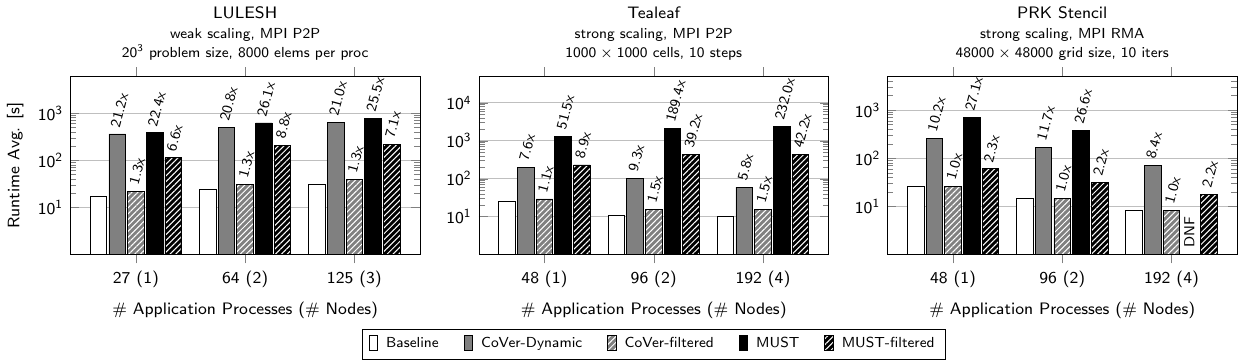}
  \caption{Overhead analysis results incl. factor over baseline, logarithmic time axis (DNF: Did not finish, deadlock after 1h).}
  \label{fig:overhead_plot}
\end{figure*}
This is not an issue for CoVer-Dynamic, as the analysis does not interfere with the parallel programming model at all.
For example, MUST intercepts MPI calls using the PMPI interface, and internally uses MPI calls to access additional information regarding the program state,
whereas CoVer-Dynamic adds its own callbacks, and has no knowledge of MPI at all apart from the provided contract annotations.

Further, the tight integration is useful for analysing MPI code, but hinders extension to further parallel programming models.
This is exemplified by the OpenSHMEM evaluation we performed; as MUST simply does not support it, it cannot report any errors.
While there are tools that can check for OpenSHMEM data races such as RMASanitizer \cite{schwitanskiRMASanitizerGeneralizedRuntime2024},
they too are limited to a specific set of known programming models.

The contract-based approach of CoVer(-Dynamic) allows for easy extension to further programming models, as demonstrated here using OpenSHMEM.
As all information required for analysis is encoded into the contract annotations,
these extensions can be simply performed by adding the corresponding rules to the necessary API functions.

\subsection{Overhead Analysis}
\label{sec:overhead_analysis}

For the overhead analysis we ran the proxy applications LULESH \cite{karlinLULESH20Updates2013}, a \cpp shock hydrodynamics model using nonblocking point-to-point MPI communication,
TeaLeaf \cite{TeaLeaf2025}, a \cpp heat conduction mini-app using collective and non-blocking point-to-point communication, and PRK Stencil \cite{vanderwijngaartParallelResearchKernels2014},
which is a C implementation of a simple stencil kernel using MPI RMA communication.

The experiments were run on compute nodes with two Intel Xeon Platinum 9242 CPUs (96 cores), 384 GB of main memory, and an InfiniBand interconnect.

Each proxy application was run using CoVer-Dynamic and MUST, both with full instrumentation as well as using filtered instrumentation, and these tests were repeated five times.
We repeated the experiment on one, two and four nodes for TeaLeaf and Stencil, and one, two and three for LULESH as LULESH only accepts cube numbers as the number of ranks.

However, MUST has issues running deadlock detection at the same time as filtered instrumentation,
which in turn causes MUST to deadlock itself. Thus, we disabled deadlock detection for the filtered MUST execution.
Further, MUST consistently deadlocked when running Stencil without filtering on three nodes; this configuration was marked as did-not-finish (DNF).
Finally, MUST requires additional tool processes to run (using threads is possible as well, but the execution timed out after one hour).
Thus, it uses more compute resources than the baseline and CoVer-Dynamic executions to run.

The results are plotted in \cref{fig:overhead_plot} (logarithmic).
All tools cause significant slowdown for the proxy app executions.
However, for CoVer-Dynamic the filtering can practically eliminate the induced overhead, as the instrumentation of \emph{every} memory location is the primary cause of the slowdown.
For MUST, the filtering is also very effective. But since MUST communicates a lot between ranks and tool processes, the effect is not as pronounced.
This can be demonstrated on the results for PRK Stencil, where CoVer-Dynamic had negligible overhead.
As CoVer did not report a (false-positive) error, no memory instrumentation is performed. Even then, MUST still caused high runtime overhead due to the communication between ranks.
Additionally, in contrast to the original testing of the static-dynamic coupling approach in \cite{orajiCouplingStaticDynamic2025} which only uses data race detection,
we are enabling most MUST features.

The MUST runtime can be improved by increasing the number of tool processes, at the cost of higher resource consumption on the compute cluster.
Additionally, MUST can run additional analyses compared to CoVer-Dynamic, such as deadlock detection (though disabled for the filtered run as mentioned),
remote data race detection for MPI RMA, and parameter checks, which may cause higher runtime impact.

Still, even without static filtering, CoVer-Dynamic finishes significantly faster on all tested configurations.
For both TeaLeaf and Stencil, CoVer-Dynamic is at least 2$\times$ faster than MUST, and 20\% faster when running LULESH.

CoVer-Dynamic has increased overhead for two nodes on TeaLeaf, which decreases again for four.
Considering the unusually high overhead MUST induces, it is safe to say that TeaLeaf, in particular, does not run well on our cluster configuration,
and can thus be considered an outlier.

Both CoVer-Dynamic and MUST reported no false positives for any of the proxy applications tested.
This is the main reason for the development of the dynamic contract analysis, as CoVer did indeed report many false positives for LULESH and TeaLeaf.
PRK Stencil is the only code where none were reported. Stencil is the smallest code, and thus does not use many complex code structures that could trip up CoVer's static analysis.

\section{Related Work}
\label{sec:related_work}

Dynamic correctness checking is possible using tools such as MUST \cite{hilbrichMUSTScalableApproach2010}, RMASanitizer \cite{schwitanskiRMASanitizerGeneralizedRuntime2024}, or ITAC \cite{ohlyAutomatedMPICorrectness2007}.
Of these, RMASanitizer is the only one to support parallel programming models other than MPI.
All of these tools work by instrumenting the code and intercepting the API calls \textquote{known to be relevant} to error detection at runtime,
then checking for erroneous behavior.
However, in contrast to CoVer-Dynamic, these tools are limited to parallel programming models and APIs that are known \emph{ahead of time}.
It is not possible to extend these tools for different parallel programming models or newer API calls without modifying the tool itself.

Static tools other than CoVer \cite{orajiVerifyingMPIAPI2026} include the SPMD IR \cite{burakSPMDIRUnifying2025} and CIVL \cite{siegelCIVLConcurrencyIntermediate2015} among others.
CoVer and the SPMD IR rely on classical data-flow analysis to detect programming errors, while CIVL uses a model checking approach with symbolic execution.
All of these tools support multiple parallel programming models.
However, apart from CoVer they still require tool modification to allow for additional programming models.
The issue with static tooling in general is the high risk of false positive results,
often induced by complex code structures and the difficulties of static pointer alias analysis.

\section{Conclusion}
\label{sec:conclusion}

In HPC, distributed-memory computing is facilitated by low-level parallel programming models, whose usage can be error-prone.
To that end, correctness tools such as CoVer~\cite{orajiVerifyingMPIAPI2026} and MUST~\cite{hilbrichMPIRuntimeError2013} help developers write and maintain bug-free codes by exposing (nondeterministic) errors such as data races, deadlocks, or resource lifetime issues.

CoVer enables model-agnostic correctness checking through a contract language that allows users to specify API usage rules without developing new analysis tools.
This design makes CoVer extensible to other (emerging) parallel programming models.
However, its static analysis limits precision regarding, e.g., data race analysis.

In this work, we introduced CoVer-Dynamic, a dynamic analysis extension for CoVer's existing contract framework.
By integrating dynamic checking, CoVer-Dynamic improves verification accuracy while avoiding duplicated tooling effort.
Dynamic analysis reduces false positives and enables detection of errors that are difficult to resolve statically, such as those involving pointer aliasing or indirect control flow.

Our evaluation compares CoVer-Dynamic against the state-of-the-art tool MUST as well as the original CoVer tool.
We show that CoVer-Dynamic achieves high classification accuracy across standardized MPI and OpenSHMEM correctness benchmarks.
CoVer-Dynamic consistently reports zero false positives and reaches a classification accuracy of up to 95\%. 
Regarding runtime overhead, we measured up to 21$\times$ overhead compared to the baseline in worst-case scenarios due to required memory instrumentation.
Nevertheless, the overhead is consistently lower than that of MUST. 
The static-dynamic tool coupling can significantly reduce this overhead up to the performance level of the baseline.
However, this may reduce detection accuracy as shown previously~\cite{orajiCouplingStaticDynamic2025} and confirmed by our classification results.

Finally, with both static and dynamic analysis available for CoVer, expanding the capabilities of its contract language is the next step.
Our tests showed that for instance deadlock detection is currently not expressible using our contract language.
Thus, future work will focus on extending support for additional error classes to broaden the applicability of the tool.

\bibliographystyle{IEEEtran}
\bibliography{IEEEabrv,bibliography}

\end{document}